\documentclass{article}
\usepackage{spconf,amsmath,graphicx}
\usepackage[bookmarks=false, hidelinks]{hyperref}
\usepackage{relsize}
\usepackage{color}
\usepackage{amssymb}
\usepackage{array}
\usepackage{textcomp}
\usepackage{multirow}
\usepackage{graphicx}
\usepackage{microtype}
\usepackage{amsmath}
\usepackage{booktabs}
\usepackage{dsfont}
\usepackage[pagewise,switch]{lineno}

\title{UNSUPERVISED CONTRASTIVE LEARNING OF SOUND EVENT REPRESENTATIONS}
\name{Eduardo Fonseca$^{1}\sthanks{Equal contribution.}$,
      Diego Ortego$^{2}\footnotemark[1]$,
      Kevin McGuinness$^{2}$, 
      Noel E. O'Connor$^{2}$,
      Xavier Serra$^{1}$
      }

\address{$^1$Music Technology Group, Universitat Pompeu Fabra, Barcelona \{eduardo.fonseca\}@upf.edu\\          
        $^2$ Insight Centre for Data Analytics, Dublin City University (DCU) \{diego.ortego\}@insight-centre.org \\
 }

\begin{document}
\ninept
\maketitle
\begin{abstract}
Self-supervised representation learning can mitigate the limitations in recognition tasks with few manually labeled data but abundant unlabeled data---a common scenario in sound event research.
In this work, we explore unsupervised contrastive learning as a way to learn sound event representations.
To this end, we propose to use the pretext task of contrasting differently augmented views of sound events.
The views are computed primarily via mixing of training examples with unrelated backgrounds, followed by other data augmentations.
We analyze the main components of our method via ablation experiments.
We evaluate the learned representations using linear evaluation, and in two in-domain downstream sound event classification tasks, namely, using limited manually labeled data, and using noisy labeled data.
Our results suggest that unsupervised contrastive pre-training can mitigate the impact of data scarcity and increase robustness against noisy labels, outperforming supervised baselines.
\end{abstract}
%
\vspace{-1mm}
\section{Introduction}
\label{sec:intro}
\vspace{-1mm}
Sound event recognition (SER) has been traditionally framed as a supervised learning problem, that is, relying on annotated datasets.
The two largest labeled SER datasets, AudioSet \cite{gemmeke2017audio} and the recently released FSD50K \cite{fonseca2020fsd50k}, are instrumental resources for SER, yet they have several shortcomings. 
AudioSet provides a massive amount of content but the official release does not include waveforms, and the labelling in some classes is less precise.\footnote{\url{https://research.google.com/audioset/dataset/index.html}}
By contrast, FSD50K consists of open-licensed audio curated with a more thorough labeling process, but the data amount is more limited.
Common to both datasets is the tremendous effort needed to collect the human annotations.

Alternatives to conventional supervised learning include the paradigms of semi-supervised learning \cite{elizalde2017approach}, few-shot learning \cite{cheng2019multi}, learning from noisy labels \cite{fonseca2020addressing,Fonseca2019model}, or self-supervision \cite{jansen2018unsupervised}.
Among these, self-supervision is appealing as it is the only one able to leverage large amounts of unlabeled data without external supervision.
Sources of unlabeled data for SER include websites such as Flickr or Freesound, which host substantial amounts of open-licensed audio(visual) material with high diversity of everyday sounds.
While content in these websites is typically accompanied by metadata, it is usually too sparse for a meaningful mapping to a sound event label set (in the case of Flickr \cite{Fonseca2019audio}), or sometimes the user-provided metadata can be an underrepresentation of the actual acoustic content (in the case of a portion of Freesound).
In these cases, self-supervision allows leveraging this unlabeled audio content in order to learn useful audio representations without prior manual labelling or metadata.


Self-supervised learning is a learning paradigm in which representations are learned by training networks on \textit{auxiliary} or \textit{pretext} tasks that do not require explicit labels.
The central idea is that by solving the pretext task, the network is able to learn useful low-dimensional representations from large amounts of unlabeled data.
These representations can then be used for downstream tasks, for example where only few data, or poorly labeled data, are available.

Self-supervised learning has shown great promise in computer vision \cite{2020_ICML_SimCLR, 2020_CVPR_MoCo}, and interesting results in the audio domain \cite{jansen2018unsupervised,tagliasacchi2020pre}.
One of the first works in self-supervised sound event representation learning is \cite{jansen2018unsupervised}, adopting a triplet loss-based training by creating anchor-positive pairs via simple audio transformations, e.g., adding noise or mixing examples.
In \cite{cartwright2019tricycle}, the pretext task consists of predicting the long-term temporal structure of continuous recordings captured with an acoustic sensor network.
Recently, \cite{tagliasacchi2020pre} proposes two pretext tasks, namely, estimating the time distance between pairs of audio segments, and reconstructing a spectrogram patch from past and future patches.
In the last few years, self-supervised learning methods using \textit{contrastive} losses have gained increasing attention, not only for images \cite{2020_ICML_SimCLR,2020_CVPR_MoCo}, but also for speech \cite{oord2018representation,nandan2020language,kharitonov2020data}, and sound events \cite{shimada2020metric}. 
In the context of contrastive learning, a recent trend is to learn representations by contrasting different versions or \textit{views} of the same data example, computed via data augmentation.
After achieving state-of-the-art in image recognition \cite{2020_ICML_SimCLR}, this approach has been successfully applied for speech recognition \cite{nandan2020language,kharitonov2020data}.

In this paper, we propose to learn sound event representations using the pretext task of contrasting differently augmented views of sound events.
The different views are computed primarily via mixing of training examples with unrelated background examples, followed by other data augmentations.
We evaluate the learned representations using linear evaluation, and in two downstream sound event classification tasks (in the same domain as the pretext task), namely, using limited manually annotated data, and using noisy labeled data.
Our results suggest that unsupervised contrastive pre-training can mitigate the impact of data scarcity and increase robustness against noisy labels.
Specifically, when training a linear classifier on top of the pre-trained embedding, we recover most of the supervised performance.
In addition, fine-tuning a model initialized with the pre-trained weights outperforms supervised baselines.
These results show that our method is able to learn useful audio representations even with more limited resources (data and compute) than other previous works (e.g., \cite{jansen2018unsupervised,cartwright2019tricycle,tagliasacchi2020pre}).
To our knowledge, this is the first work conducting contrastive sound event representation learning by maximizing the similarity between differently augmented views.
Our contributions are \textit{i)} a new framework for learning sound event representations via data-augmenting contrastive learning, and \textit{ii)} an empirical evaluation of the proposed method through extensive experiments. Code is available.\footnote{\url{https://github.com/edufonseca/uclser20}\label{repo}}



\section{Method}
\label{sec:method}
\begin{figure}
  \centering{}\includegraphics[width=1\columnwidth]{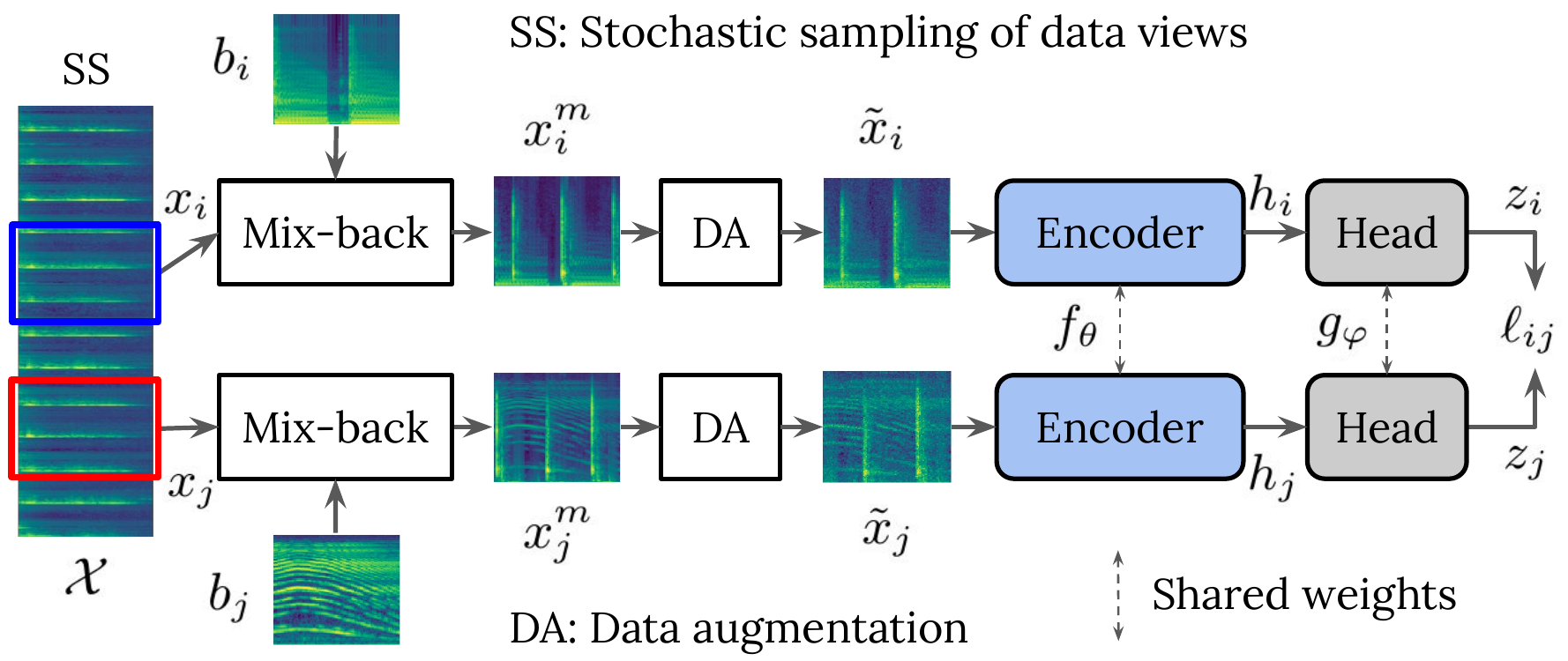}
  \vspace{-5mm}
  \caption{Contrastive learning of sound event representations overview.}
  \label{fig:Method-overview.}
  \vspace{-3mm}
\end{figure}
We seek learning sound event representations from unlabeled data using
self-supervision, i.e., exploiting a pretext task whose \textit{labels} are inferred
from an unlabeled dataset. To do so, we create pairs of correlated
views (denoted as positive examples) via different augmentations of
a single sound event example. Then, their corresponding embedding representations
are compared using a contrastive loss \cite{2020_ICML_SimCLR,2020_ICMLW_SpeechSimCLR,2020_ECCV_ContrastAudioVis}
that pulls together representations of positive examples, while pushing
apart those of negative ones (i.e., unrelated examples). Our hypothesis
is that discriminative sound event representations can emerge by solving
this task. Figure \ref{fig:Method-overview.} illustrates the main
components of the proposed method, which is inspired by the recent
SimCLR \cite{2020_ICML_SimCLR}.
Next we explain the main components.
Implementation details and hyper-parameter choices can be inspected in the released code.\textsuperscript{\ref{repo}}

\paragraph*{Stochastic sampling of data views.}
The incoming training examples to our framework are log-mel spectrograms of audio clips.
From each training example, $\mathcal{X}$, we sample two views, which we call time-frequency (TF) \textit{patches}.
These patches, $x_{i}\in\mathcal{X}$ and $x_{j}\in\mathcal{X}$ are selected randomly over the length of the clip spectrogam. 
Sec. \ref{subsec:DrawingPositives} analyzes the benefits of this stochastic sampling over other alternatives.

\paragraph*{Mix-back.}
The first operation that we apply to each incoming patch is what we call \textit{mix-back}.
It consists in \textit{mix}ing the incoming patch $x_{i}$ with a \textit{back}ground patch, $b_{i}$, as follows:
\begin{equation}
x_{i}^{m}=\left(1-\lambda\right)x_{i}+\lambda\left[E\left(x_{i}\right)/E\left(b_{i}\right)\right]b_{i},\label{eqn:mixback}
\end{equation}
where $\lambda\sim\mathcal{U}\left(0,\alpha\right)$, $\mathcal{U}$ is a uniform distribution, $\alpha\in\left[0,1\right]$ is the mixing hyper-parameter (typically small), and $E\left(\cdot\right)$
denotes the energy of a given patch. 
A similar approach is used in \cite{jansen2018unsupervised} to create examples for triplet loss-based training.
The energy adjustment of Eq. \ref{eqn:mixback} ensures that $x_{i}$ is always dominant over $b_{i}$, even if $E\left(b_{i}\right)>>E\left(x_{i}\right)$, thereby preventing aggressive transformations that may make the pretext task too difficult.
Before Eq. \ref{eqn:mixback}, patches are transformed to linear scale (inversion of the log in the log-mel) to allow energy-wise compensation, after which mix-back is applied, and then the output, $x_{i}^{m}$, is transformed back to log scale.
Background patches $b$ are randomly drawn from the training set (excluding the input clip $\mathcal{X}$), hence they are out-of-batch in the vast majority of cases.
Recent work \cite{2020_arXiv_GoodViews} shows that useful representations arise if data views share as little information as possible, while preserving relevant semantic information that keeps the predictive power for related downstream tasks.
This is our motivation to use mix-back: \textit{i)} shared information across positives is decreased by mixing $x_{i}$ and $x_{j}$ with different backgrounds, and \textit{ii)} semantic information is preserved due to sound transparency (i.e., a mixture of two sound events inherits the classes of the constituents) and the fact that the positive patch is always predominant in the mixture.
Mix-back can be understood as a data augmentation, but we separate it from the others as it involves two input patches.

\paragraph*{Stochastic Data Augmentation.}
We adopt data augmentation techniques (DAs) directly computable over TF patches (rather than waveforms), and that are simple for on-the-fly computation, thus favouring speed rather than acoustical/mathematical correctness.
We consider DAs both from computer vision and audio literature: random resized cropping (RRC), random time/frequency shifts, compression, specAugment \cite{2019_INTERSPEECH_SpecAugment}, Gaussian noise addition, and Gaussian blurring. 
All augmentations are stochastic as their hyper-parameters are randomly sampled from a distribution for each patch. 
The final augmentation policy adopted consists of sequentially applying RRC, compression and Gaussian noise addition (see Sec. \ref{ssec:DA_exp} for other policies).
These DAs transform $x_{i}^{m}$ into the input patch $\tilde{x}_{i}$ for the encoder network.

\paragraph*{Encoder Network.}

We use a CNN based network $f_{\theta}$ to extract the embedding $h_{i}=f_{\theta}\left(\tilde{x}_{i}\right)$ from the augmented patch $\tilde{x}_{i}$, where $h_{i}$ is the embedding right before the final fully-connected classification layer and $\theta$ are its parameters. 
Once the contrastive learning is over and the encoder is trained, the representation $h_{i}$ can be used for downstream tasks.

\paragraph*{Projection Head.}
Following \cite{2020_ICML_SimCLR}, a simple projection network $g_{\varphi}$ with parameters $\varphi$ maps $h_{i}$ to the final L2-normalized low-dimensional representation $z_{i}$ where the contrastive
loss is applied.
Our head consists of an MLP with one hidden layer, batch-normalization, and a ReLU non-linearity.
Note that the projection head is only used during contrastive learning, i.e., once the training is over, only the trained encoder is used for downstream tasks.

\paragraph*{Contrastive Loss.}
The contrastive loss adopted for a positive pair of examples, $x_{i}$ and $x_{j}$, is the \textit{NT-Xent} loss \cite{2020_ICML_SimCLR}:
\begin{equation}
\ell_{ij}=-\log\frac{\exp\left(z_{i}\cdot z_{j}/\tau\right)}{\sum_{v=1}^{2N}\mathds{1}_{v\neq i}\exp\left(z_{i}\cdot z_{v}/\tau\right)},\label{eq:ContrastiveLoss}
\end{equation}
where $z_{j}$ is the representation for the patch $x_{j}$, $\tau>0$ is a temperature scaling, $\mathds{1}_{v\neq i}\in\left\{ 0,1\right\}$ is an indicator function that returns 1 if $v\neq i$, and $N$ is the batch size.
Note that the generation of two views for each example extends the mini-batch from $N$ to $2N$.
Therefore, every patch has a single positive pair and $2N-2$ negative pairs.
Minimizing Eq. \ref{eq:ContrastiveLoss} during training adjusts the parameters $\theta$ and $\varphi$ to maximize the numerator (i.e., maximize agreement between positives) while simultaneously minimizing the denominator, thus forcing similar views to neighboring representations and dissimilar ones to non-neighboring ones. 
The DA operations over each view of one example are two different instantiations of the same family of transformations.
However, the embeddings $z_{i}$ and $z_{j}$ are obtained with a single instantiation of the encoder and
projection head (see \textit{shared weights} in Figure \ref{fig:Method-overview.}).


\section{Experimental Setup}
\label{sec:setup}
\subsection{Dataset}
\label{ssec:dataset}

We use the FSDnoisy18k dataset \cite{Fonseca2019learning}, containing 42.5h of Freesound audio distributed across 20 classes drawn from the AudioSet Ontology \cite{gemmeke2017audio}.
The dataset includes a small \textit{clean} training set (1,772 clips / 2.4h), a larger \textit{noisy} train set (15,813 clips / 38.8h), and a test set (947 clips/ 1.4h).
Labels in the clean and test sets are manually-labelled with \textit{Freesound Annotator} \cite{Fonseca2017freesound}, whereas labels in the noisy set are inferred automatically from metadata, hence featuring real-world label noise.
The dataset is singly- and weakly-labeled, and clips are of variable-length in range [0.3, 30]s.
We avoid larger vocabulary datasets, e.g., AudioSet \cite{gemmeke2017audio}, due to the computationally intensive contrastive learning experiments conducted---intractable under our compute resources. 
FSDnoisy18k has a smaller vocabulary, while featuring a relatively large amount of per-class training data compared to other datasets (\cite{salamon2014dataset,piczak2015esc,fonseca2018general,Fonseca2019learning,cartwright2019sonyc}). 
The limited dataset scope implies, however, that the learned representations are not transferable to unrelated datasets or downstream tasks, forcing to conduct an \textit{in-domain} evaluation.
FSDnoisy18k allows evaluation of learned representations on two real-world scenarios: using a small clean set, and a larger noisy set.

\subsection{Learning Pipeline}
\label{ssec:pipeline}
Our pipeline consists of two stages: \textit{i)} unsupervised contrastive learning of a low-dimensional representation (Sec. \ref{sec:contrastive}), and \textit{ii)} evaluation of the representation using supervised tasks (Sec. \ref{sec:eval}).
In both stages, 
incoming audio is transformed to 96-band log-mel spectrograms, and to deal with variable-length clips, we use TF patches of 1s (shorter clips are replicated; longer clips are trimmed in several patches). 
We use three networks: ResNet-18 \cite{2016_CVPR_ResNet}, and a VGG-like and a CRNN similar to those in \cite{fonseca2020fsd50k}, commonly used for SER tasks (details can be inspected in the code\textsuperscript{\ref{repo}}).
Models are always trained using SGD with momentum 0.9 and weight decay $10^{-4}$, using a batch size of 128 and shuffling examples between epochs.
We always train on the noisy set and validate on the clean set, except in one of the downstream tasks in Sec. \ref{ssec:eval_downstream}.
For the contrastive learning experiments (Sec. \ref{sec:contrastive}), we follow the approach of Sec. \ref{sec:method}.
Models are trained for 500 epochs, with initial learning rate of 0.03, divided by 10 in epochs 325 and 425. 
For the supervised learning experiments (Sec. \ref{sec:eval}), during training we randomly sample a single patch when clips are longer than 1s.
Models are trained for 200 epochs to minimize categorical cross-entropy, reducing the learning rate in epochs 80 and 160.
The initial learning rate is 0.1 when training from scratch, and 0.01 in case of using unsupervised pre-trained weights in order to constrain the learning process.


\subsection{Evaluation}
\label{ssec:eval}

To quantify the quality of the learned representations, we follow three approaches: a variation of the standard $k$-Nearest Neighbour (\textbf{kNN}) \textbf{evaluation} in \cite{2018_CVPR_InstanceDis}, and the standard approaches of using \textbf{linear probes} \cite{2020_ICML_SimCLR} and  \textbf{fine-tuning a model end-to-end}.
For the \textbf{kNN evaluation}, we estimate the representation $z$ for each validation or test patch and compare it against every other patch in the given set via cosine similarity. The prediction for every patch is, then, obtained by majority voting rule across the $k$ neighbouring labels, where $k=200$ as in \cite{2018_CVPR_InstanceDis,2019_ICML_AND}. 
In turn, clip-level predictions are obtained by majority voting of patch-level ones.
This evaluation is used for fast contrastive learning experimentation as no additional training is involved.
Unlike kNN evaluation, \textbf{linear probes} and \textbf{end-to-end fine-tuning} procedures involve further training and passing patches through an entire model to produce prediction probabilities.
The former involves training an additional linear classifier on top of the pre-trained unsupervised embedding, whereas the latter fine-tunes the model on a given downstream task after initializing it with the pre-trained weights.
For both procedures, patch-level predictions are averaged per-class across all patches in a clip to obtain clip-level predictions.
Common to all evaluation methods, once clip-level predictions are gathered for a given set, overall accuracies are computed on a per-class basis, then averaged with equal weight across all classes to yield the performance shown in Secs. \ref{sec:contrastive} and \ref{sec:eval}.
For linear probes and end-to-end fine-tuning, we report test accuracy provided by the best validation accuracy model.
However, learning curves for the contrastive learning experiments using kNN eval were found to be relatively noisy, such that \textit{best} accuracy is not always representative of the overall quality of the training process.
Therefore, we decided to use the \textit{average} validation accuracy across the last 50 epochs, as top performing model checkpoints appear at the end of the training.
Finally, results by the three methods above are compared against supervised baselines, where metrics are computed following the same procedure as in end-to-end fine-tuning.
Hereafter, we shall refer to validation as \textit{val}.
\vspace{-2mm}

\section{Ablation Study}
\label{sec:contrastive}

We report ablation experiments to study some blocks of our proposed method.
For each block, we always start from the best configuration.

\subsection{Sampling TF Patches \label{subsec:DrawingPositives}}
Table \ref{tab:samplePatch} shows results when patches are randomly sampled along the clip, and when patches are sampled deterministically, separated by a sampling distance $d$ (time frames).
Stochastically sampling patches provides superior performance.
We study the impact of progressively increasing $d$ (might be bounded due to clip length) between a first randomly sampled patch $x_i$ and a second patch  $x_j$. 
The higher the distance, the better the representation learned. 
In particular, for $d<101$ both patches overlap and these cases underperform no overlapping ones, where performance saturates.
This aligns with recent results in computer vision, where increasing the distance between image crops is beneficial only up to some values, after which performance decreases due to little semantic content shared between views \cite{2020_arXiv_GoodViews}. 
\begin{table}[t]
\vspace{-4mm}
\caption{kNN val accuracy for several ways of sampling TF patches.}
\vspace{+1mm}
\centering
\begin{tabular}{lc|lc}
\toprule
\textbf{Sampling method}  & \textbf{kNN} & \textbf{Sampling method}  & \textbf{kNN} \\
\midrule
Sampling at random     & \textbf{70.1}         & $d = 125$             & 67.9     \\
$d = 0$ (same patch)            & 51.1         & $d = 200$             & 69.9     \\
$d = 25$                        & 61.5         & $d = 300$             & 68.5     \\
$d = 75$                        & 65.1         & $d = 400$             & 69.7  \\
\bottomrule
\end{tabular}
\label{tab:samplePatch}
\vspace{-6mm}
\end{table}

\subsection{Mix-back}
\label{ssec:mixback_exp}
Table \ref{tab:mixback} (left side) shows results for mix-back for best $\alpha \in \left\lbrace 0.02,0.05,0.1, 0.2\right\rbrace$.
It can be seen that using mix-back helps considerably, and adjusting the energy is also beneficial. 
The latter means that the foreground patch is always dominant over the background patch, thus preventing potentially aggressive transforms.
The optimal $\alpha$ values are small, which indicates that a light mixture is preferred.
We observe that lightly adding background from another patch (i.e., mix-back) is much more beneficial than adding artificial white noise (see first and second rows from Table \ref{tab:mixback} (right side)).
These results suggest that mixing with natural backgrounds from real audio signals is suitable for contrastive representation learning. 
\begin{table}[t]
\vspace{-4mm}
\caption{kNN val accuracy for several mix-back and data augmentation (DA) settings.}
\vspace{+1mm}
\centering
\begin{tabular}{lc|lc}
\toprule
\textbf{Mix-back setting ($\alpha$)}  & \textbf{kNN} & \textbf{DA policy}  & \textbf{kNN} \\
\midrule
w/ $E$ adjustment (0.05)     & \textbf{70.1}  &  RRC + comp + noise            & \textbf{70.1}     \\
w/o $E$ adjustment  (0.02)   & 66.2         & RRC + comp                      & 69.6     \\
w/o mix-back                  & 63.3        & RRC + specAugment                 & 70.0  \\ 
                             &              & RRC                              & 69.0     \\
                             &              & specAugment \cite{2019_INTERSPEECH_SpecAugment} & 68.0  \\
                             &              & w/o DA                        & 60.1     \\
\bottomrule
\end{tabular}
\label{tab:mixback}
\vspace{-6mm}
\end{table}

\vspace{-2mm}
\subsection{Data Augmentation}
\label{ssec:DA_exp}

Table \ref{tab:mixback} (right side) lists results for several DA policies.
Each row represents the best result after sweeping the corresponding DA parameters.
We started by exploring DAs applied individually (from bottom to top).
The top DA (applied individually) is random resized cropping (RRC).
In particular, the optimal RRC found applies a mild cropping (instead of a harsh one), which can be seen as a small stretch in time and frequency (which also involves a small frequency transposition).
This RRC slightly outperforms specAugment \cite{2019_INTERSPEECH_SpecAugment}, which has been successfully used for contrastive learning of speech embeddings \cite{nandan2020language}, and which also works well for sound events.
Then, we explored DA compositions based on the mentioned RRC.
We found that, to a lesser extent, compression (via spectrogram image contrast adjustment) and Gaussian noise addition also improve the learned representation.
We adopt this DA policy for all experiments reported in Secs. \ref{sec:contrastive} and \ref{sec:eval} (top row in Table \ref{tab:mixback}).
A subsequent more thorough DAs exploration revealed promising results by composing RRC and specAugment, yielding almost top results (70.0).
It seems possible that a more exhaustive (and costly) exploration of the DA compositions may lead to better results, e.g., complementing RRC and specAugment with other soft DAs such as compression or Gaussian noise addition.
Nonetheless, this should be done carefully. 
For example, in our experiments we have seen that the ordering of the DAs matter, and joining individually-tuned DAs can be suboptimal as different DAs in a composition affect each other.
\vspace{-2mm}





\subsection{Encoder and Temperature}
\label{ssec:encoder_temp}
For encoder architectures, we explore ResNet-18, VGG-like and CRNN, obtaining 70.1, 67.7 and 67.1 val kNN accuracies, respectively. This results show that higher capacity (ResNet-18) is better for contrastive learning, which accords with \cite{2020_ICML_SimCLR}.
We also experiment with the temperature $\tau$ of the contrastive loss function, which has been found to be a relevant parameter \cite{2020_CVPR_MoCo}.
In particular we sweep $\tau \in \left\lbrace 0.1, 0.2, 0.3, 0.4\right\rbrace$, obtaining the kNN accuracy of 68.9, 70.1, 68.9 and 67, respectively.
Results show the framework's sensitivity to $\tau$, which we also find to be highly dependent on the projection head configuration.  
\vspace{-2mm}

\subsection{Discussion}
In general, we observe that the framework is sensitive to hyper-parameter changes, and that the various settings of each block affect each other, thus requiring extensive experimentation for appropriate tuning.
From Secs. \ref{subsec:DrawingPositives} and \ref{ssec:mixback_exp} we observe that overlapping patches when drawing positives is detrimental, and lightly adding natural backgrounds is beneficial.
This could indicate that the original positive examples sometimes share time-frequency patterns that could be used to lower the loss of Eq. \ref{eq:ContrastiveLoss}, but that hinder the learning of useful representations.
These undesired patterns 
are denoted as \textit{shortcuts} in computer vision \cite{minderer2020automatic}.
Examples of shortcuts in audio self-supervised learning include e.g., recording gear, room acoustics or background noise.
FSDnoisy18k is based on Freesound audio, which in turn is composed of audio contributed by users.
Clips coming from the same user are likely to share some of these patterns, which has been shown to have an impact in supervised sound event tagging \cite{fonseca2020fsd50k}.
We hypothesize that this could be a source of shortcuts in our setting, which is being mitigated by stochastic sampling of positive patches and mix-back. 
We intend to further investigate this in future work.
Finally, previous works report advantages of using larger batch sizes as a way to provide more negative examples, which facilitates convergence (e.g., sizes up to 8k are used in \cite{2020_ICML_SimCLR}).
Our compute resources do not support such large sizes, hence all our experiments are conducted with a batch size of 128 (commonly used for supervised learning).
It is conceivable that using larger batches yields even better results than those reported here.

\section{Evaluation of Learned Representations}
\label{sec:eval}

\subsection{Baseline Systems and Linear Evaluation}
\label{ssec:eval_linear}
\begin{table}[t]
\vspace{-4mm}
\caption{Test accuracy for linear probes evaluation (second column), and for two downstream sound event
classification tasks: a larger noisy set and a small clean set for training. *This is also the supervised baseline to compare with linear evaluation. p-t = pre-trained.}
\vspace{+1mm}
\centering
\begin{tabular}{l@{}|c|cc|cc@{}}
\toprule 
\textbf{Model}  &  \textbf{Linear} & \multicolumn{2}{c|}{\textbf{Larger noisy set}} & \multicolumn{2}{c}{\textbf{Small clean set}}  \\
\midrule 
(weights in M)        & -    & random*    & p-t   & random    & p-t  \\
\midrule 
ResNet-18 (11)    & \textbf{74.3}  & 65.4      & \textbf{78.2}         & 56.5  & \textbf{77.9}       \\
VGG-like (0.3)    & 70.0          & 70.6      & \textbf{72.8}         & 61.1  & \textbf{72.3}       \\
CRNN  (1)        & 64.4          & 72.0      & \textbf{74.2}         & 58.7      & \textbf{69.1}   \\
\bottomrule
\end{tabular}
\label{tab:FinetuneEval}
\vspace{-6mm}
\end{table}
Table \ref{tab:FinetuneEval} presents the test accuracies for the linear probes evaluation and the supervised baselines (second and third columns).
The supervised CRNN and VGG-like perform similarly, while ResNet-18 performs worse, similarly as in \cite{fonseca2020fsd50k} with FSD50K.
In our case, this could be due to the capacity of the model (the largest by far), which may lead to overfitting of the smaller dataset (and the noisy labels).
In linear evaluation of the contrastive weights, however, ResNet-18 is the top performing system, which accords with the kNN evaluation (Sec. \ref{ssec:encoder_temp}) and with findings in \cite{2020_ICML_SimCLR}.
Thus, with ResNet-18 the supervised baseline is exceeded by a considerable margin, whereas with VGG-like and CRNN most of the supervised performance is recovered (99\% and 89\%, respectively).

\vspace{-1mm}
\subsection{In-Domain Downstream Prediction Tasks}
\label{ssec:eval_downstream}
We conduct an in-domain evaluation (see Sec. \ref{ssec:dataset}), similarly to \cite{cartwright2019tricycle}, for the two downstream sound event classification tasks posed by FSDnoisy18k: training on the larger set of noisy labels, 
and on the small set of clean data (in this case, 15\% of the clean set is kept for validation).
For each task, we compare \textit{i)} a supervised baseline trained from scratch, with \textit{ii)} fine-tuning the network initialized with unsupervised pre-trained weights---these correspond respectively to the columns \textit{random} and \textit{p-t} (pre-trained) in Table \ref{tab:FinetuneEval}.
These experiments aim at measuring benefits with respect to training from scratch in noisy- and small-data regimes.
Table \ref{tab:FinetuneEval} shows that unsupervised contrastive pre-training brings great benefits, achieving better results than training from scratch in both tasks and across all network architectures considered.
For both tasks, using ResNet-18 yields top accuracy in the pre-trained setup, and the lowest accuracy when trained from scratch. 
This suggests that the performance attainable with ResNet-18 supervised from scratch is limited, potentially by limited data and/or label quality.
In contrast, unsupervised contrastive pre-training seems to alleviate these problems, leveraging ResNet-18's capacity and yielding superior performance.
Greater improvements are observed in the ``smaller clean'' task, where, interestingly, the pre-trained performance shows little degradation with respect to that of the ``larger noisy'' task (despite having far fewer examples, see Sec. 3).
Specifically, for ResNet-18, pre-trained performance decreases from 78.2 to 77.9, while training from scratch yields a substantial accuracy drop (65.4 to 56.5).
A possible explanation for the similar pre-trained performance across tasks may be that, in the ``smaller clean'' task, the pre-trained model is fine-tuned with unseen clean data (albeit small).
However, in the ``larger noisy'' task, the model is fine-tuned with the same data previously used for unsupervised contrastive learning (and the supervision provided is affected by label noise).
\vspace{-1mm}





\section{Conclusion}
\label{sec:conclusion}
We present a framework for unsupervised contrastive learning of sound event representations, based on maximizing the similarity between differently augmented views of the same log-mel spectrogram.
Via ablation experiments, we show that appropriately tuning the compound of positive patch sampling, mix-back, and data augmentation is vital for successful representation learning.
The evaluation on in-domain sound event classification tasks suggests that unsupervised contrastive pre-training can mitigate the impact of data scarcity, and increase robustness against noisy labels as recently found in supervised image classification \cite{2019_ICML_PreTraininglabelNoise}.
Future work includes exploring time-domain DAs, and using larger-vocabulary datasets to learn general-purpose representations transferable to out-of-domain tasks.



\section{Acknowledgments}
This work is partially supported by Science Foundation Ireland (SFI) under grant number SFI/15/SIRG/3283 and by the Young European Research University Network under a 2020 mobility award. Eduardo Fonseca is partially supported by a Google Faculty Research Award 2018. The authors are grateful for the GPUs donated by NVIDIA.

\bibliographystyle{IEEEbib}
\bibliography{strings,refs}

\end{document}